# Machine studies for the development of storage cells at the ANKE facility of COSY

K. Grigoryev[a,b], F. Rathmann[b], R. Engels[b], A. Kacharava[b], F. Klehr[c], B. Lorentz[b], S. Martin[b], M. Mikirtychyants[a,b], D. Prasuhn[b], J. Sarkadi[b], H. Seyfarth[b], H.J. Stein[b], H. Ströher[b], A. Vasilyev[a]

*a) Petersburg Nuclear Physics Institute, Gatchina, Russia*

*b) Institut für Kernphysik, Jülich Center for Hadron Physics, Forschungszentrum Jülich GmbH, Germany*

*c) Zentralabteilung Technologie, Forschungszentrum Jülich GmbH, Germany*

**Abstract**

We present a measurement of the transverse intensity distributions of the COSY proton beam at the target interaction point at ANKE at the injection energy of 45 MeV, and after acceleration at 2.65 GeV. At 2.65 GeV, the machine acceptance was determined as well. From the intensity distributions the beam size is determined, and together with the measured machine acceptance, the dimensions of a storage cell for the double-polarized experiments with the polarized internal gas target at the ANKE spectrometer are specified. An optimum storage cell for the ANKE experiments should have dimensions of 15 mm $_{\text{vertical}}$ × 20 mm $_{\text{horizontal}}$ × 390 mm $_{\text{longitudinal}}$, whereby a luminosity of about $2.5 \cdot 10^{29}$ cm$^{-2}$s$^{-1}$ with beams of $10^{10}$ particles stored in COSY could be reached.

## 1 Introduction

The demand for high luminosity internal targets for nuclear and hadron physics experiments at storage rings has led to remarkable developments during the past decade, in particular for unpolarized targets. The cluster targets for Hydrogen (H) and Deuterium (D) used for instance at COSY[1] [1] are routinely operated at densities of $5 \cdot 10^{14}$ to $10^{15}$ cm$^{-2}$ [2]. Together with the high intensity, unpolarized and polarized proton and deuteron beams of COSY, luminosities of $5 \cdot 10^{30}$ to $10^{31}$ cm$^{-2}$s$^{-1}$ are reached [3].

The present investigation is motivated by the need to also provide high luminosities for the foreseen double-polarized studies of few-nucleon systems at COSY [4]. The best polarized atomic beam sources (ABS[2]) for H and D today reach fluxes near $10^{17}$ s$^{-1}$. The densities for instance reported with a polarized H gas jet target at the EDDA experiment at COSY were $2 \cdot 10^{11}$ cm$^{-2}$ [5]. Storage cells can be efficiently used to enhance the target density of a free atomic jet from an ABS by about two orders in magnitude [6], and such a storage cell should be available for the measurements with the polarized internal gas target (PIT[3] [7]) at the ANKE[4] spectrometer [8].

The determination of the beam size at the location of the target at different energies is crucial for the design of the storage cell. One usually uses a beam-profile monitor to obtain information about the beam size [9]. At present, such a device is not available at COSY and therefore we developed a system that facilitates the measurement of the beam size by moving the edge of a scraper through the stored beam. Such a system has the advantage that the size can be directly determined at the location of the ANKE target, while the use of beam-profile monitors involves additional uncertainties from lattice calculations, which relate the beam size at the beam-profile monitor to the one at the target.

---

[1] **Co**oler **Sy**nchrotron, Forschungszentrum Jülich GmbH, Germany

[2] **A**tomic **B**eam **S**ource

[3] **P**olarized **I**nternal gas **T**arget

[4] **A**pparatus for studies of **N**ucleon and **K**aon **E**jectiles

The paper is organized as follows. In Section 2, we give a short description of the method employed to obtain the horizontal and vertical beam size from the measured intensity distributions by means of a movable beam scraper. In Section 3, we briefly present some considerations concerning storage cells. In Section 4, the experimental setup used for the measurement is described, including a characterization of the COSY lattice in terms of β functions and dispersion. The measurement of the beam size carried out at 45 MeV (injection) is presented in Section 5.1, and for 2.65 GeV in Section 5.2. Normalized beam emittances at 45 MeV and 2.65 GeV are compared in Section 5.3. In Section 6 the determination of the machine acceptance is discussed. An estimate of the achievable luminosity based on the obtained results and the storage cell considerations of Section 3 is presented in Section 7, followed by the conclusions (Section 8).

## 2 Determination of the beam size by a movable scraper

It is reasonable to assume that the stored beam has a two-dimensional Gaussian distribution in transverse phase space. The betatron amplitude distribution in e.g. the y-y' phase space is then given by [10], [11]

$$\rho_\beta(y) = \frac{N_0}{\sigma_y^2} \cdot y \cdot \exp\left(-\frac{y^2}{2\sigma_y^2}\right). \qquad (1)$$

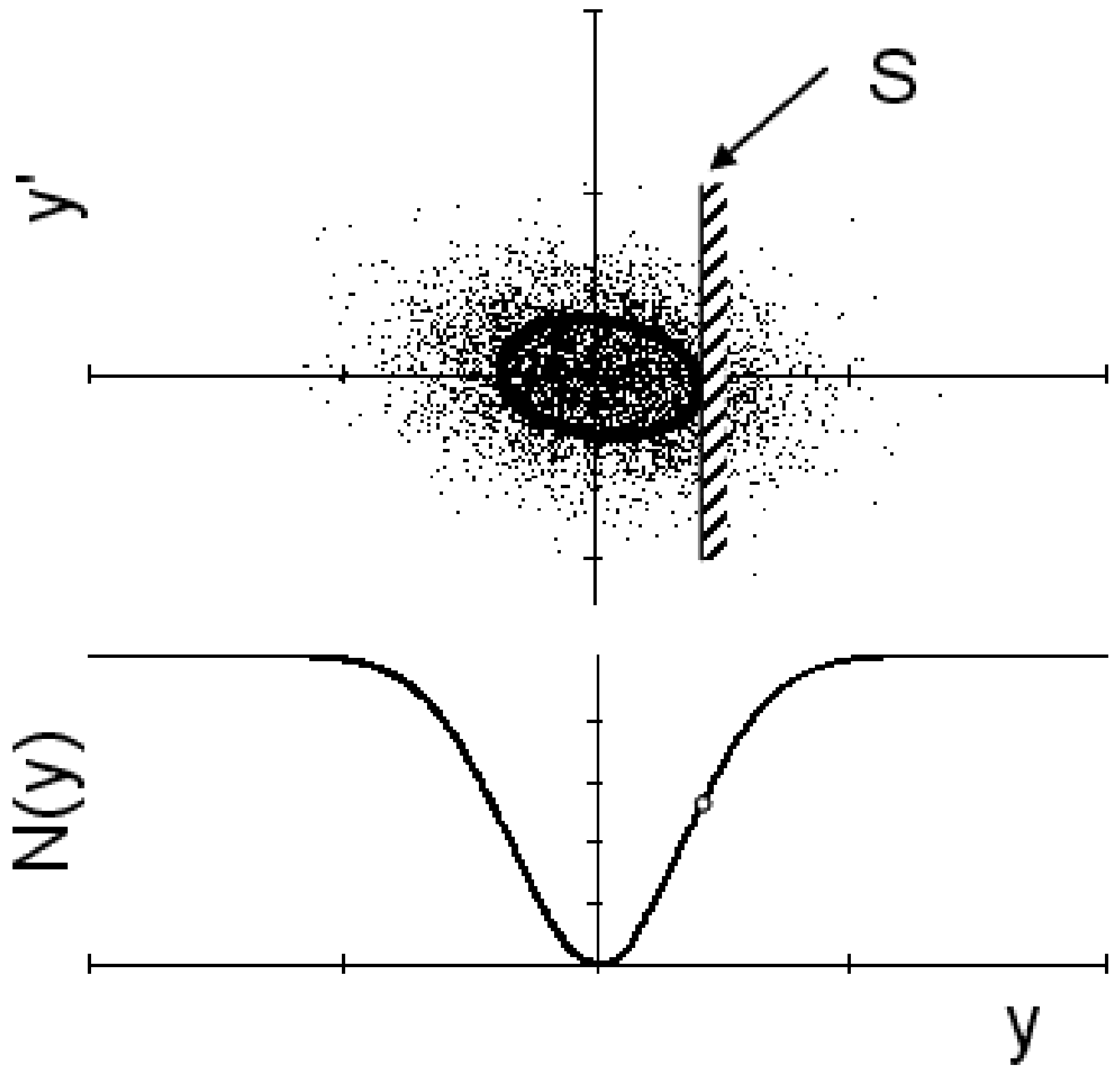

**Fig. 1.** y-y' phase space distribution of the beam at the location of a movable scraper S (top panel). In case the edge of the scraper is far away the beam intensity is not affected. Moving the aperture into the beam to a position y, removes all particles with betatron amplitudes larger than the distance of the aperture to the beam center, i.e. all particles that are outside the indicated phase space ellipse. The measured beam current (empty circle in bottom panel) is given by N(y) (Eq. (2)).

In the absence of coupling, a properly aligned scraper moving along the y (x) direction will only remove particles from the y-y' (x-x') phase space with betatron amplitudes larger than the distance from the beam center to the scraper edge, as illustrated in Fig. 1. With a scraper at a location y, the beam intensity can be written as

$$N(y) = \int_0^{y-\mu_y} \rho_\beta(y)\cdot dy = N_0 \cdot \left[1-\exp\left(-\frac{(y-\mu_y)^2}{2\sigma_y^{\,2}}\right)\right]. \tag{2}$$

This function can be fitted to the measured beam intensity distribution, and the position of the beam center $\mu_y$ and the beam width $\sigma_y$ can be determined.

## 3 Storage cell considerations

A storage cell consists of a feeding and a beam tube, as shown in Fig. 2. The gas density distribution in a storage cell has a maximum at the location of feeding tube.

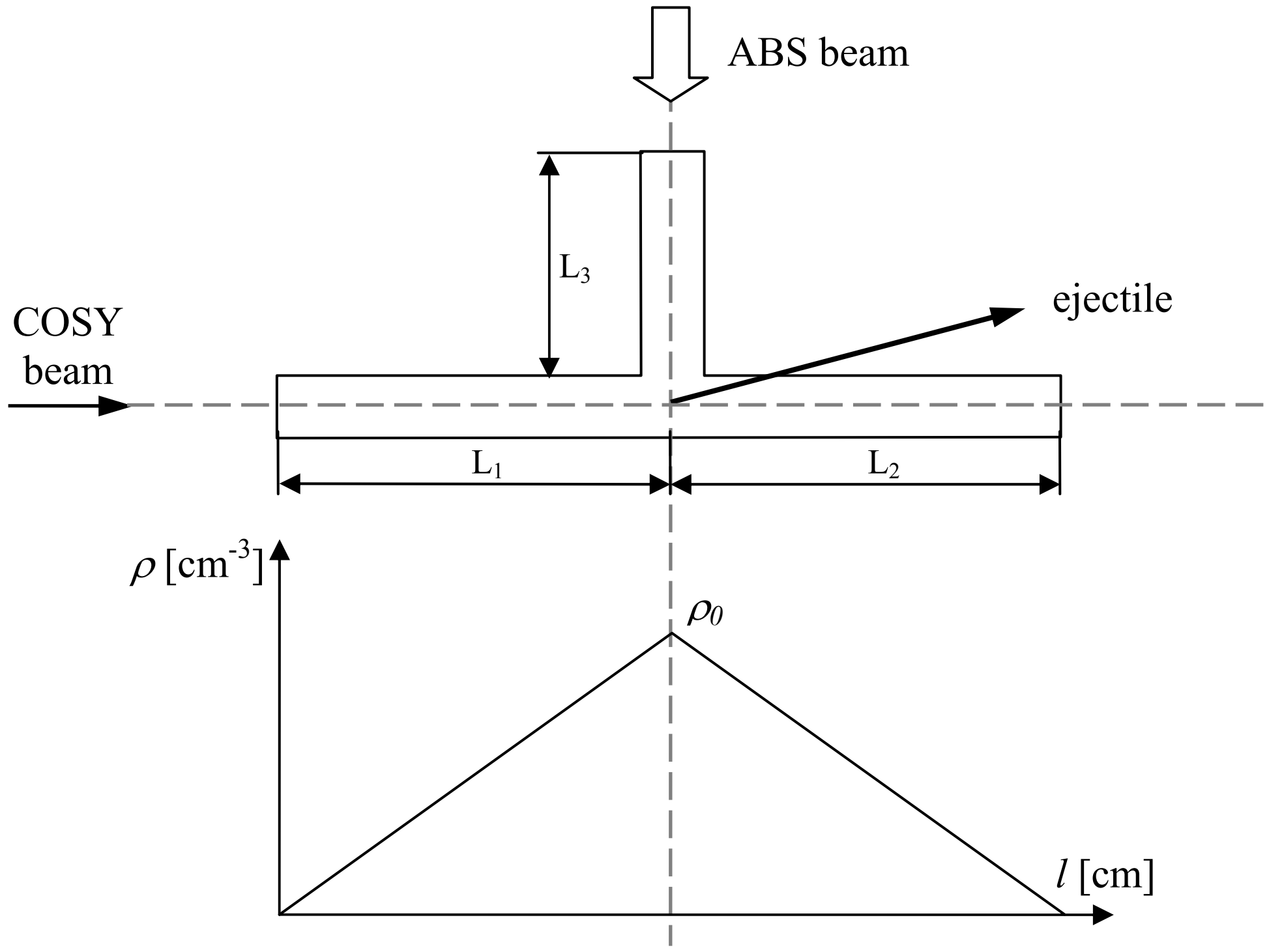

**Fig. 2.** Sketch of the storage cell (top panel). The atomic beam from the ABS is injected into the cell through the feeding tube. The gas density distribution along the beam tube in the storage cell is shown in the bottom panel.

The central density is given by the balance equation

$$\rho_0 = I_{ABS}/C_{tot}\,, \tag{3}$$

where $I_{ABS}$ [s$^{-1}$] is the polarized beam intensity from the source. The total conductance [cm$^3$/s] of a cell consisting of three tubes of circular cross section [12] is given by

$$C_{tot} = 3.81\cdot 10^3 \cdot \sqrt{\frac{T}{M}}\left\{\sum_{i=1}^{3}\frac{D_i^{\,3}}{L_i + 1.33\cdot D_i}\right\}, \tag{4}$$

where $T$ [K] denotes the temperature of the walls of the cell, $M$ [amu] is the molar mass, $D_i$ [cm] and $L_i$ [cm] are diameter and length of the tubes ($i$ = 1,2,3), respectively. Assuming a linear decrease of the gas density from the center of the storage cell to the exits of the beam tube, the areal target density in atoms/cm$^2$ is given by

$$d_t = \frac{1}{2} \cdot (L_1 + L_2) \cdot \rho_0 \propto \frac{1}{D_1^3}, \tag{5}$$

where $D_1$ (= $D_2$) denotes the beam-tube diameter.

The critical parameter to achieve a high target thickness is thus the diameter of the beam tube; therefore, one tries to make it as small as possible. On the other hand, the transverse storage-cell dimensions are limited by the lateral dimensions of the beam at injection and by its transverse motion during acceleration. Based on the known the size of the ABS beam [13], the measured COSY beam size at ANKE, together with the measured machine acceptance (presented in Sections 5 and 6), it is straightforward to specify storage cell dimensions which provide reaction rates near the optimum in the experiments (presented in Section 7).

## 4 Experimental setup

The measurements were carried out at the ANKE target chamber with an unpolarized proton beam at 45 MeV (injection) and, after acceleration, at 2.65 GeV. All measurements used coasting beams, i.e., neither electron nor stochastic cooling was employed, and the COSY cavities were not used either. Thus, the mean energy loss of the stored beam was not compensated. The ANKE hydrogen cluster target [1] was utilized to provide a heating of the beam during the present investigations at 2.65 GeV similar to that when the atomic beam is injected into the storage cell. The cluster target provided an areal density of about $10^{14}$ atoms/cm$^2$.

### 4.1 Beam scraper system

For the measurements, a dedicated support system has been constructed (Fig. 3). It consists of two three-axis XYZ manipulators, which support a frame with a set of diaphragms, as shown in Fig. 4. The setup is operated in the stray field of the spectrometer magnet D2. In order to avoid magnetic forces, the frame and the diaphragms were made from aluminum. The inner dimensions of the large diaphragm were 25 mm $_{\text{ver}}$ × 50 mm $_{\text{hor}}$, i.e., substantially larger than the expected beam size to keep all injected particles in the beam. A set of small diaphragms (10 mm $_{\text{ver}}$ × 30 mm $_{\text{hor}}$) was installed in the support frame as well. These diaphragms were much smaller and were used to find out how many particles could remain in the beam with a storage cell of this cross section. The storage cell prototype (Fig. 4) was not used during the measurements presented here.

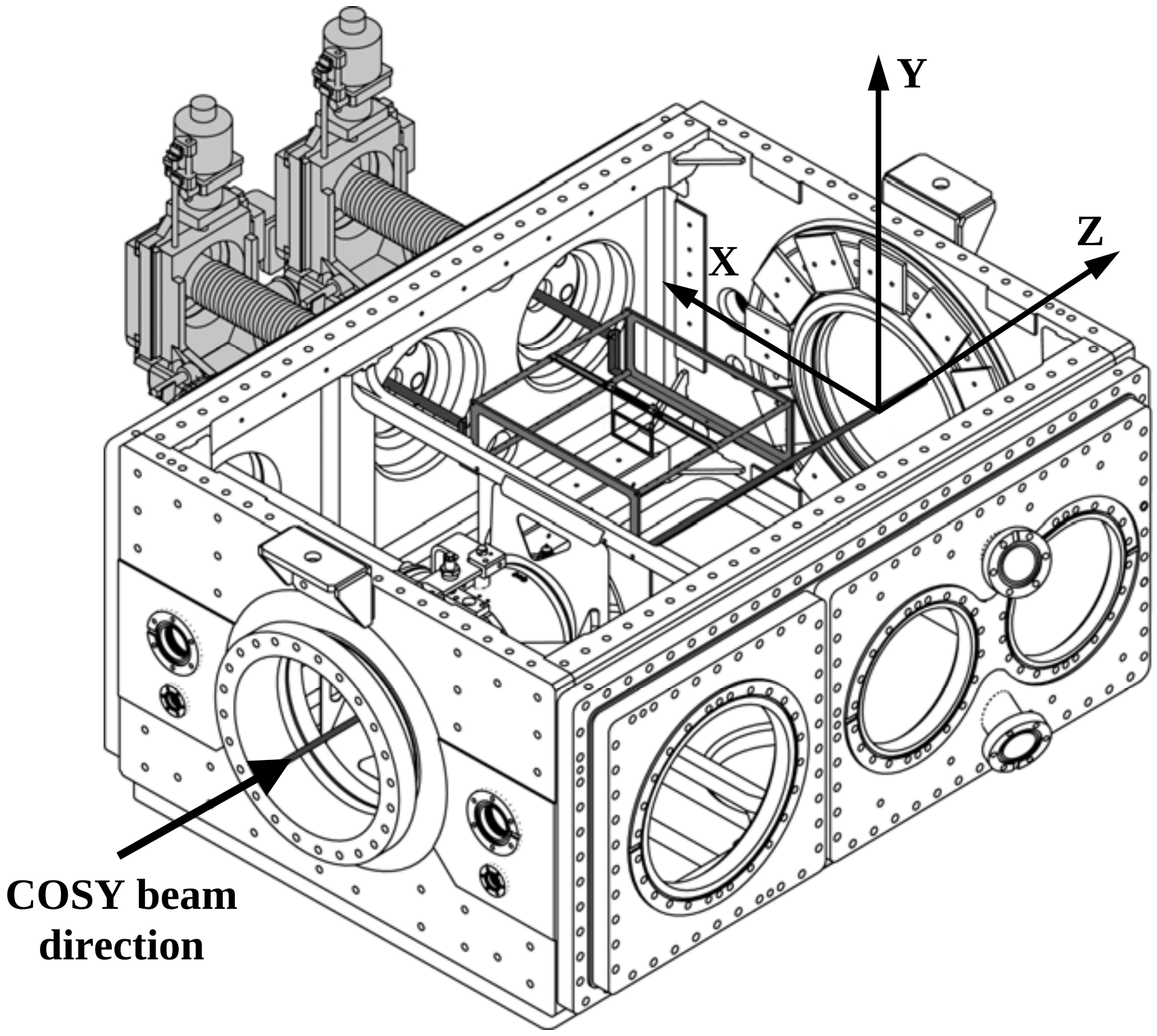

**Fig. 3.** ANKE target chamber with XYZ manipulators and support frame. The coordinate system used in the measurements is indicated as well.

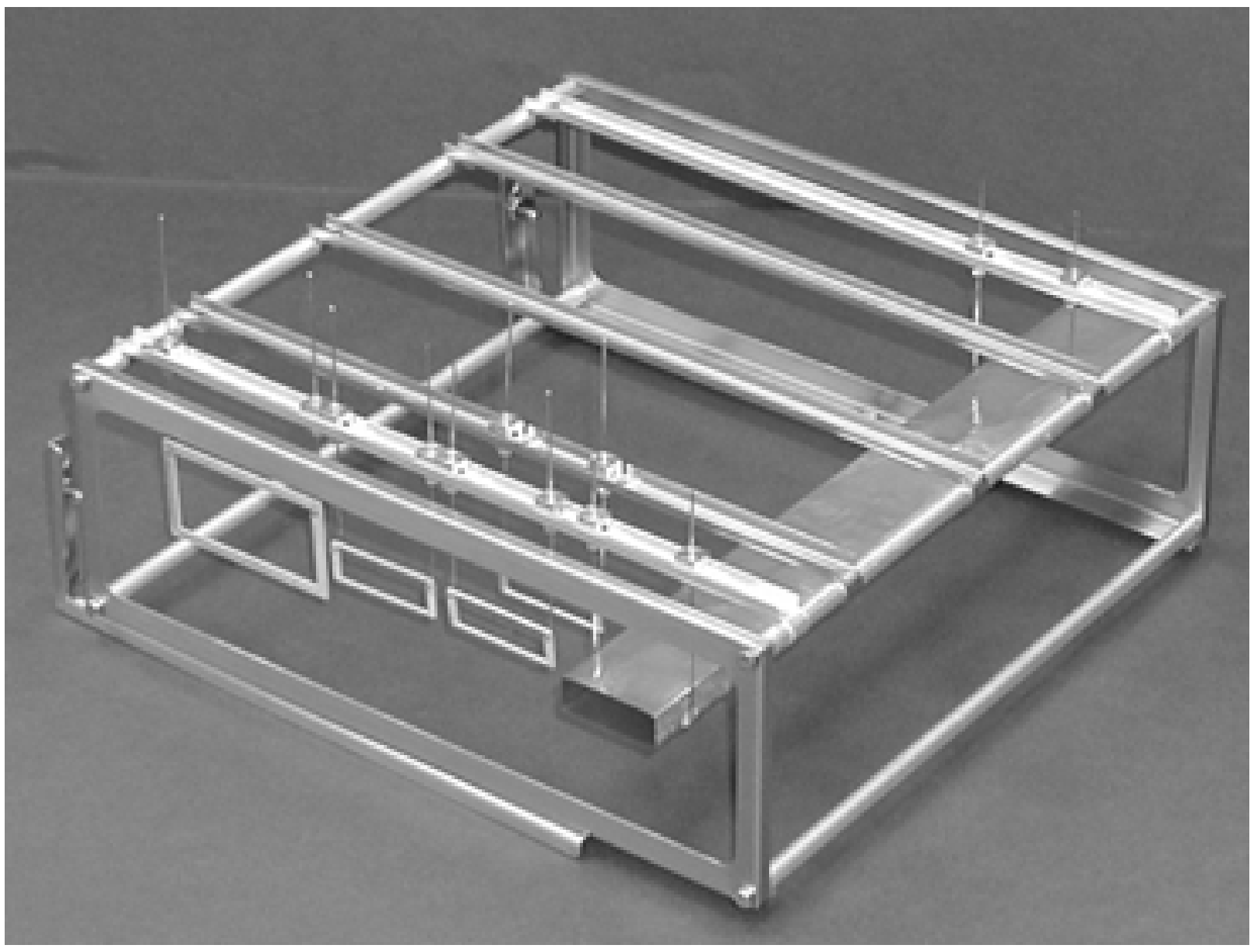

**Fig. 4.** Support frame equipped with one large and a set of small diaphragms.

The setup was mounted on a rectangular flange of the target chamber. The support frame could be moved by precise stepper motors[5] with a step size of 0.25 μm. The stepper-motor system is equipped with position sensors. The accuracy of the positions of the diaphragm, determined by the step size, was controlled by a position sensor. A computer with WinCC software[6] [14] was used for remote control of the frame position from the COSY control room. The scheme of the control system, used to operate the manipulators via a ProfiBUS[7] connection between the accelerator hall and the setup in the accelerator tunnel, is shown in Fig. 5.

[5] Vacuum Generators HPT Translator, MRXMOTX and MRXMOTZ, Producer – Vacuum Generators Ltd, UK

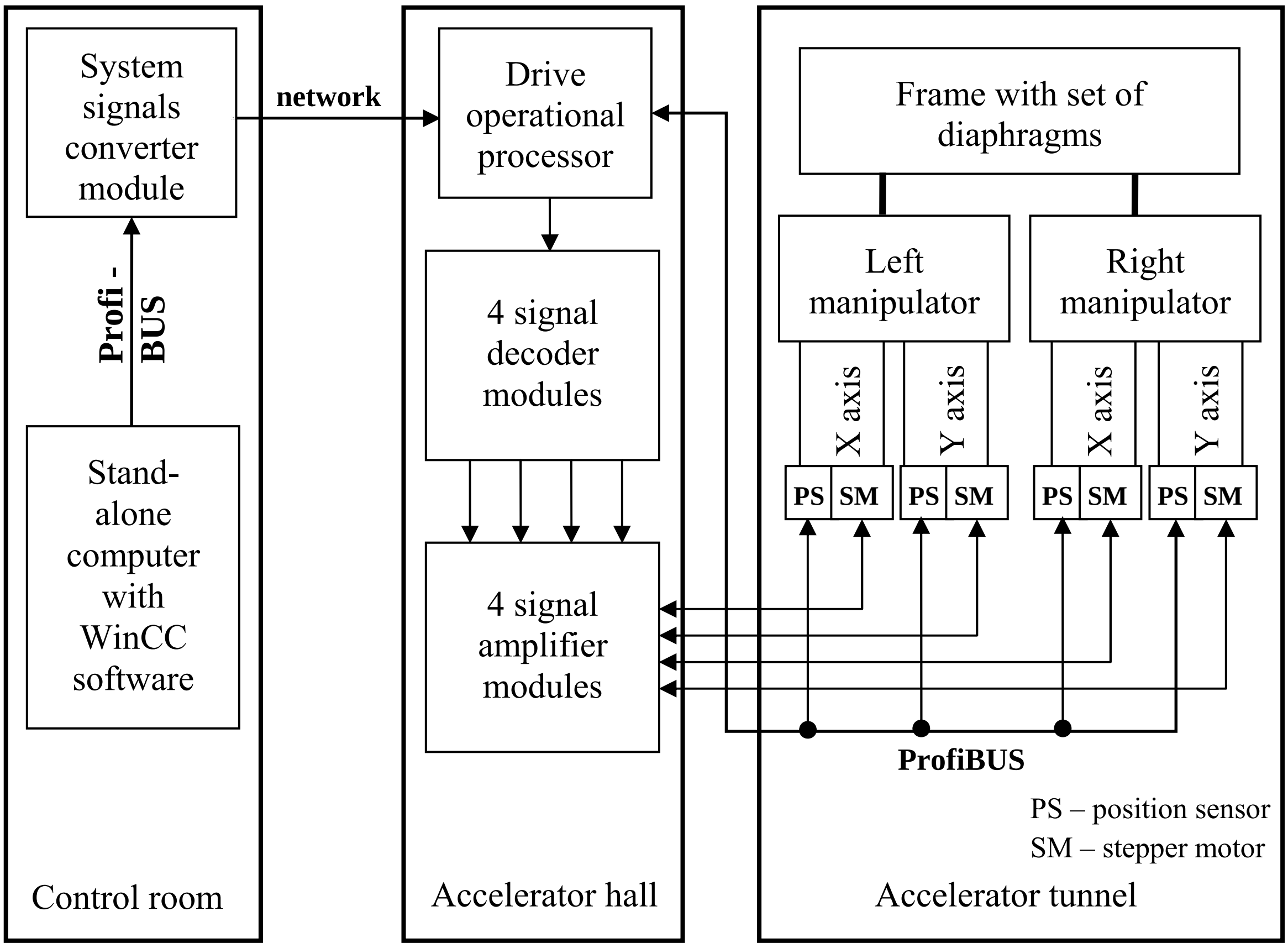

**Fig. 5.** Remote control scheme used to operate the XYZ manipulators from the COSY control room.

The bellows of the manipulators allow one to move the frame by 150 mm along the horizontal (x) axis in the chamber and by ± 12.5 mm in the up/down (y) and forward/backward (z) direction (the coordinate system is shown in Fig. 3). In addition, the frame can be completely removed from the beam. Thus, the machine could be operated for other experiments without breaking of the vacuum and dismounting the flange with the support system. For security reasons, the setup was equipped with end-switch sensors to restrict the movement of the support frame inside the vacuum chamber. These sensors were adjusted to stop further movement of the support frame in order to protect the vacuum bellows of the manipulators as well as other equipment inside the target chamber, like the Silicon tracking telescopes [15].

The alignment of the apertures for the measurements was carried out in a separate test chamber of dimensions matching those of the ANKE target chamber with a laser beam passing through the optical

[6] Windows based SCADA (Supervisory Control And Data Acquisition) system of Siemens AG, Automatisierungs- und Antriebstechnik, D-90475 Nürnberg, Germany

[7] ProfiBUS DP (Decentralized Periphery) – field bus, based on the international standard IEC 61158

center of the chamber. During this procedure, the coordinates of the centers of the various diaphragms were determined.

### 4.2 Characterization of the COSY lattice at 45 MeV and 2.65 GeV

The betatron amplitude functions $\beta_x$ ,$\beta_y$ and the horizontal dispersion $D_x$ at 45 MeV and at 2.65 GeV are presented in Fig. 6. Also indicated are the locations of cluster target and ABS beam, apertures and the extension of the storage cell. Precise values for the β functions are also given in Table 1 and Table 2. At 2.65 GeV, the horizontal dispersion at ANKE is small (below ~ 0.2 m), while in the arcs $D_x$ ~ 15 m.

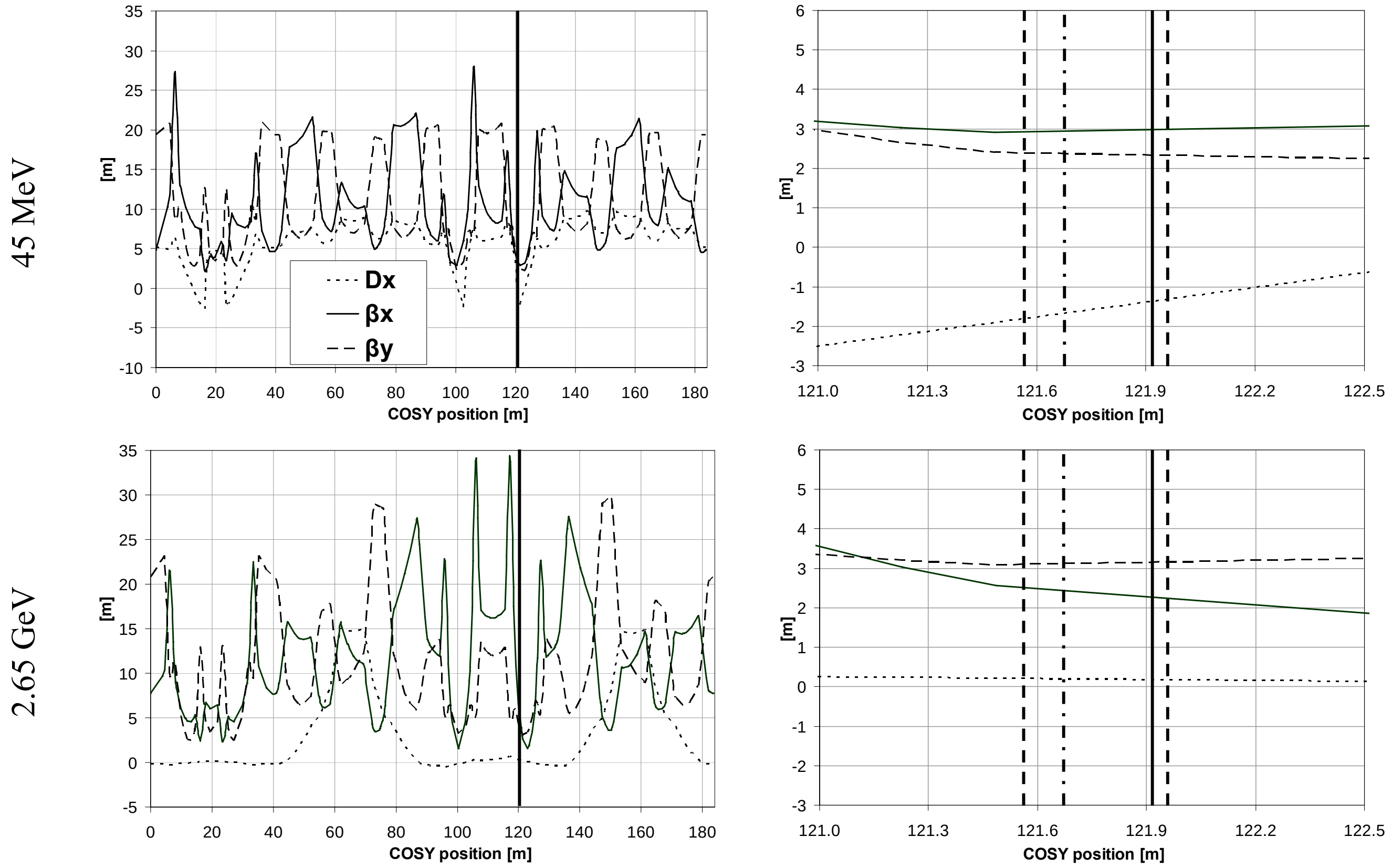

**Fig. 6.** Betatron functions $\beta_x$ ,$\beta_y$ and horizontal dispersion $D_x$ at 45 MeV (top panels) and at 2.65 GeV (bottom). The panels on the left show the lattice functions for the complete ring, those on the right on an expanded scale only the region near the ANKE target region. Left panels: The thick vertical line indicates the ANKE interaction point. Right panels: The vertical solid thick line indicates the position of the cluster target, the dash-dotted line the location of the large aperture (Fig. 4), which is the same as the location of the ABS beam. The vertical dashed lines indicate the extension of the storage cell.

A beam-current transformer (BCT) measured the current orbiting in the machine. The beam current is transformed into a voltage and readout with an oscilloscope in the COSY control room, using

$$U_{BCT} = c \cdot I_{BEAM} = c \cdot N \cdot f \cdot e\,, \qquad (6)$$

where $I_{BEAM}$ [A] is the current induced by $N$ particles orbiting in the machine with frequency $f$ [Hz], $e$ [As] is the charge of the particle, $c$ [Ohm] = 100 is the total resistance of the COSY beam-current transformer. For the measurements, the BCT was carefully adjusted to read zero when the beam was completely removed from the machine.

# 5 Beam size and beam emittance measurements

## 5.1 Beam size at 45 MeV (injection)

Prior to the beam-size measurements, the large diaphragm was centered in the ANKE target chamber and the intensity of the injected beam was compared to the situation when the support frame was completely withdrawn from the beam. It was shown that the beam intensity was not affected by the large diaphragm, i.e., the large diaphragm did not restrict the machine acceptance at injection.

During every injection into the storage ring, the large diaphragm was centered on the beam. At the end of the injection process, the diaphragm was moved sideways in horizontal (or vertical) direction. The number of protons that remain in the machine after a horizontal (or vertical shift) of the diaphragm from the center position is shown in Fig. 7. The flat part of the distribution characterizes the situation when the frame does not affect the stored beam. After each measurement, prior to the next injection, the diaphragm was returned to the center position.

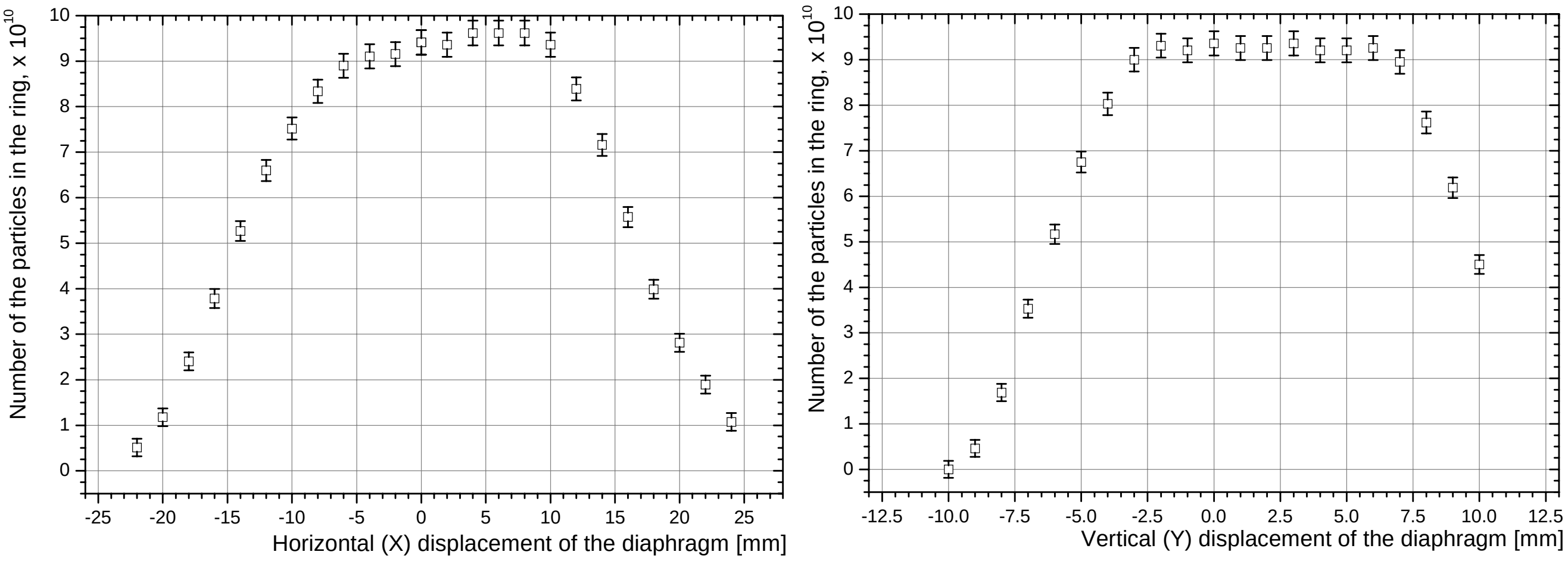

**Fig. 7**. Number of stored protons at 45 MeV in the ring as function of the horizontal (left) and vertical (right) displacement of the diaphragm from the optical center of the target chamber.

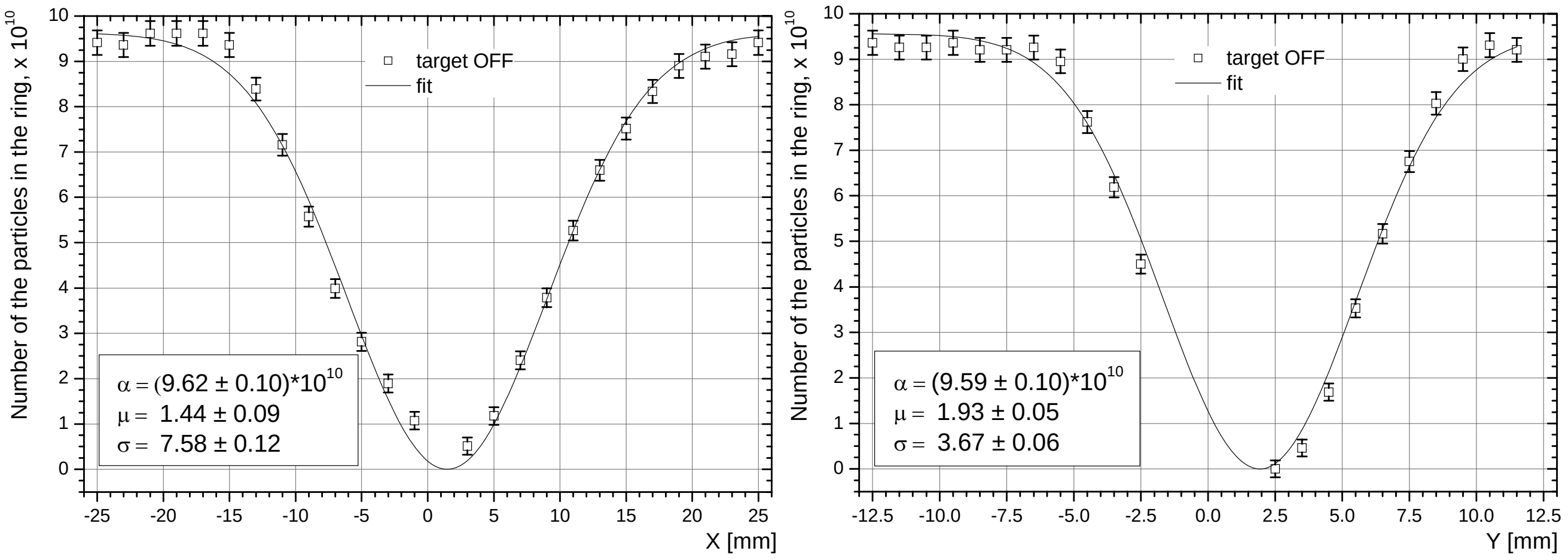

**Fig. 8.** Number of stored particles in COSY as function of horizontal (left panel) and vertical (right) position at 45 MeV at the ANKE interaction point. The results of fits using Eq. (2) are given in the insets.

When the diaphragm was horizontally (vertically) displaced by ± 25 mm (± 12.5 mm), the beam was completely lost. In order to obtain the beam intensity as function of the position using the data of Fig. 7, the positive part of the distribution, $x \geq 0$ ($y \geq 0$), is shifted to the left by half the diaphragm width (height), and the negative part, $x < 0$ ($y < 0$), is shifted to the right by the same amount, yielding the beam intensity as function of position, shown in Fig. 8.

As discussed in Section 2, the resulting beam intensity distributions, shown in Fig. 8, can be described with a Gaussian function, using Eq. (2), where $N_0$ denotes the maximum beam intensity, $\mu$ is the shift of the beam from the optical center of the target chamber, and $\sigma$ is the Gaussian width. The beam intensity was fitted using an uncertainty for the beam current of

$$\Delta \mathrm{I} = \sqrt{(0.02 \cdot \mathrm{I})^2 + \left(0.02 \cdot \max(\mathrm{I})\right)^2}\ , \qquad (7)$$

to account for possible fluctuations of the BCT offset and a scale-factor uncertainty. The resulting parameters are given in Fig. 8. The uncertainties fully account for correlations between the errors. The derived two-dimensional beam intensity distribution at injection at ANKE is shown in Fig. 9. Based on the measured horizontal and vertical beam widths and the β functions, known from the lattice calculations (see Fig. 6), we give in Table 1 the 2σ-beam emittances, usually used for proton beams [10].

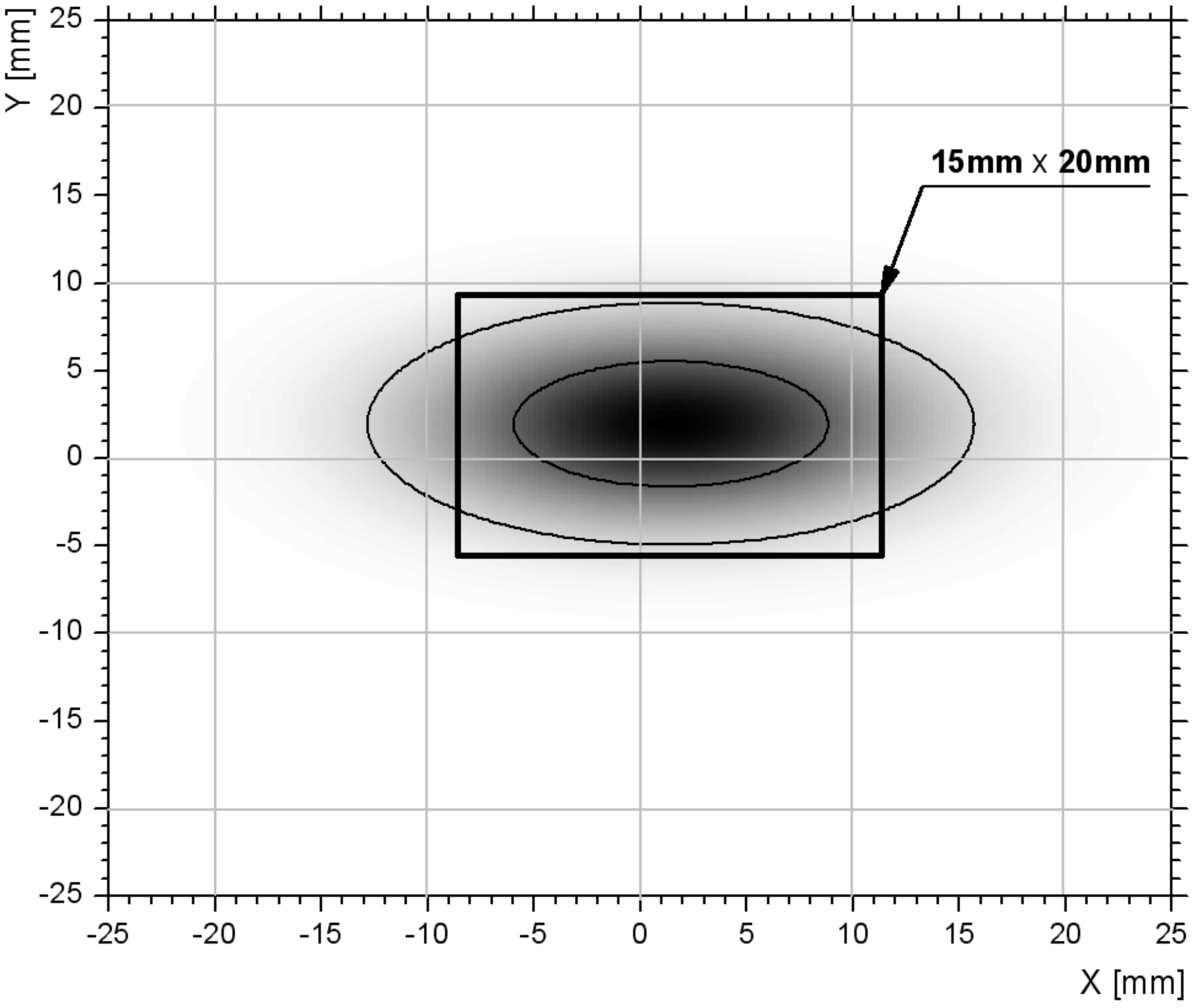

**Fig. 9.** Measured two-dimensional COSY beam intensity distribution at 45 MeV at ANKE, where the two ellipses correspond to 1σ and 2σ beam emittances. Also shown is the cross section of the storage cell (see Section 7).

**Table 1.** Parameters of the proton beam at 45 MeV. Based on the measured beam widths σ and the known β functions from the lattice calculations at the ANKE interaction point, the horizontal and vertical 2σ-beam emittance is calculated, taking into account a relative error of 10% for the β function.

| | σ | β function | Δβ/β | beam emittance $(2\sigma)^2/\beta$ |
|---|---|---|---|---|
| | [mm] | [m] | | [π mm mrad] |
| x | 7.58 ± 0.12 | 2.8 | 0.1 | 85.12 ± 8.93 |
| y | 3.67 ± 0.06 | 2.7 | 0.1 | 19.24 ± 2.02 |

## 5.2 Beam size at 2.65 GeV

After the acceleration procedure was completed, with the stored beam passing through the center of the large diaphragm, the diaphragm was moved to a new position for the measurement. Subsequently, during a cycle of ten minute duration, the number of particles in the ring was recorded every 60 seconds. Besides obtaining the size of the accelerated beam, an additional aim was to also determine the beam lifetimes for different positions of the diaphragm. (These measurements are further discussed in Section 6.) After the end of the cycle, the diaphragm was moved back to the center position for the next injection. The data were taken with and without the ANKE cluster target (target off and target on).

The intensity distributions of the beam at 2.65 GeV are obtained from a procedure similar to the one applied at 45 MeV. The results of this procedure for the horizontal and vertical distributions with and without target on flattop after 1 min are shown in Fig. 10. It was not possible to access the central region of the beam during the measurements along the vertical axis, because of the restricted motion of the XYZ manipulators and a vertical displacement of the beam of about 3 mm with respect to the center of the target chamber (Fig. 10, top panels).

A careful analysis using the error matrix was carried out, with uncertainties $\Delta I$ of the beam current $I$ using the convention given in Eq. (7). Within the derived uncertainties (indicated in Fig. 10) no change of the beam parameters is observed when the target is switched on. Therefore, a weighted average of the horizontal and vertical beam widths with and without target was taken, and the 2σ-beam emittance was determined using the known β functions (including an error of 10%). The results are listed in Table 2, indicating that the vertical beam emittance is about twice as large as the horizontal one.

**Table 2.** Parameters of the proton beam after acceleration to 2.65 GeV. Based on the measured beam widths σ and the β functions from the lattice calculations at the ANKE interaction point, the horizontal and vertical 2σ-beam emittance are calculated, taking into account a relative error of 10% for the β function.

| | σ | β function | Δβ/β | beam emittance $(2\sigma)^2/\beta$ |
|---|---|---|---|---|
| | [mm] | [m] | | [π mm mrad] |
| x | 1.19 ± 0.02 | 2.6 | 0.1 | 2.19 ± 0.23 |
| y | 1.72 ± 0.05 | 3.1 | 0.1 | 3.82 ± 0.43 |

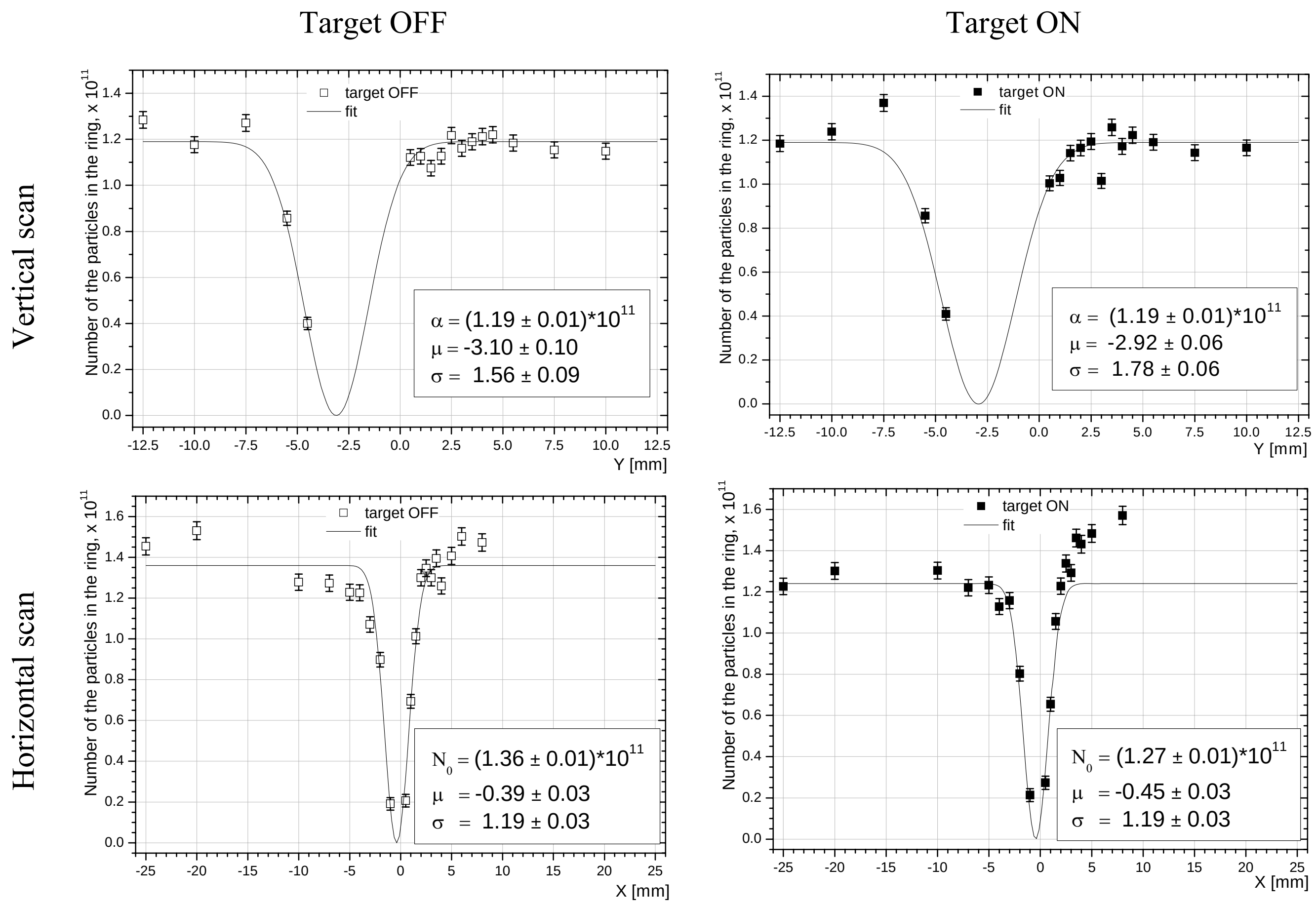

**Fig. 10.** Number of stored particles in COSY with and without target (right and left panels) as function of horizontal and vertical position (bottom and top) at 2.65 GeV at the ANKE interaction point. The results of fits using Eq. (2) are given in the insets.

### 5.3 Comparison of normalized beam emittance at different energies

The measured beam emittances at 45 MeV and at 2.65 GeV can be compared in terms of normalized emittances. The normalized emittance of a beam is a so-called adiabatic invariant, defined through

$$\varepsilon_N = \varepsilon \cdot \beta_{rel} \cdot \gamma_{rel}\,, \tag{8}$$

where $\beta_{rel}$ and $\gamma_{rel}$ are the relativistic kinematical parameters. The conversion of the obtained results from Table 1 and Table 2 yield the normalized beam emittances listed in Table 3. The reason for the reduced normalized horizontal beam emittance is most likely caused by a mismatch between frequency of the accelerating cavity and the dipole field. This leads to orbit excursions in the machine, which in turn lead to particle loss during acceleration. We attribute the surprising increase in the vertical beam emittance to the variation of machine tunes during acceleration. Near machine resonances the beam size increases, but the acceleration procedure is fast (~ 1s) so that the time spent near a machine resonance is too short for particles to be lost immediately, but long enough to increase the beam emittance.

**Table 3.** Comparison of the normalized beam emittances for 45 MeV and 2.65 GeV.

| | 45 MeV | | 2.65 GeV | |
|---|---|---|---|---|
| | measured 2σ beam emittance [π mm mrad] | $\varepsilon_N$ [π mm mrad] | measured 2σ beam emittance [π mm mrad] | $\varepsilon_N$ [π mm mrad] |
| x | 85.12 ± 8.93 | 26.68 ± 2.80 | 2.19 ± 0.23 | 8.08 ± 0.85 |
| y | 19.24 ± 2.02 | 6.03 ± 0.63 | 3.82 ± 0.43 | 14.10 ± 1.59 |

## 6 Machine acceptance measurements

The determination of the beam sizes alone, described in Sections 5.1 and 5.2 is not sufficient to specify the transverse dimensions of the storage cell. These results are necessary to determine the *minimum* possible cell size, regardless of the constraints at injection energy. For a cell that can be opened at injection energy, which shall be developed at a later stage, these constraints do not apply. In this section we address the measurement of the machine acceptance measurements at 2.65 GeV to find out how *large* the cross section of the storage cell should be in order to not intercept the machine acceptance.

### 6.1 Determination of the machine acceptance with a movable scraper

The method employed to determine the machine acceptance is based on the analysis of the beam lifetimes, which were obtained together with the beam size and beam emittance measurement at 2.65 GeV (Section 5.2). A similar investigation has been carried out in conjunction with the storage cell development for the polarized target experiments at the Indiana Cooler [16]. In order to illustrate the approach, we show in Fig. 11 the y-y' phase-space distribution for the measured vertical beam emittance $\varepsilon_y = 3.8\ \pi$ mm mrad at 2.65 GeV at the ANKE interaction point. The orientation of the phase space distribution is given by the three lattice functions α, β, and γ, as discussed in ref. [10]. With $\beta_y = 3.1$ m, $\alpha_y = 0.17$, and $\gamma_y = (1+\alpha_y^2)/\beta_y$ the ellipse in Fig. 11 (left panel) corresponds to a machine acceptance $A_y = 25\ \pi$ mm mrad. When the aperture restriction, indicated by the vertical dashed bar, is located outside of the machine acceptance, the beam lifetime is not affected. Moving the aperture into the phase space ellipse of the machine, leads to a reduction of the beam lifetime, because the maximum allowed angle y' for single Coulomb scattering is reduced. From $y^2 = \varepsilon_y \cdot \beta_y$ and $y'^2 = \varepsilon_y \cdot \gamma_y$ it follows that $y' = y \cdot \sqrt{\gamma_y/\beta_y}$, and thus the maximum allowed angle y' is strictly proportional to y.

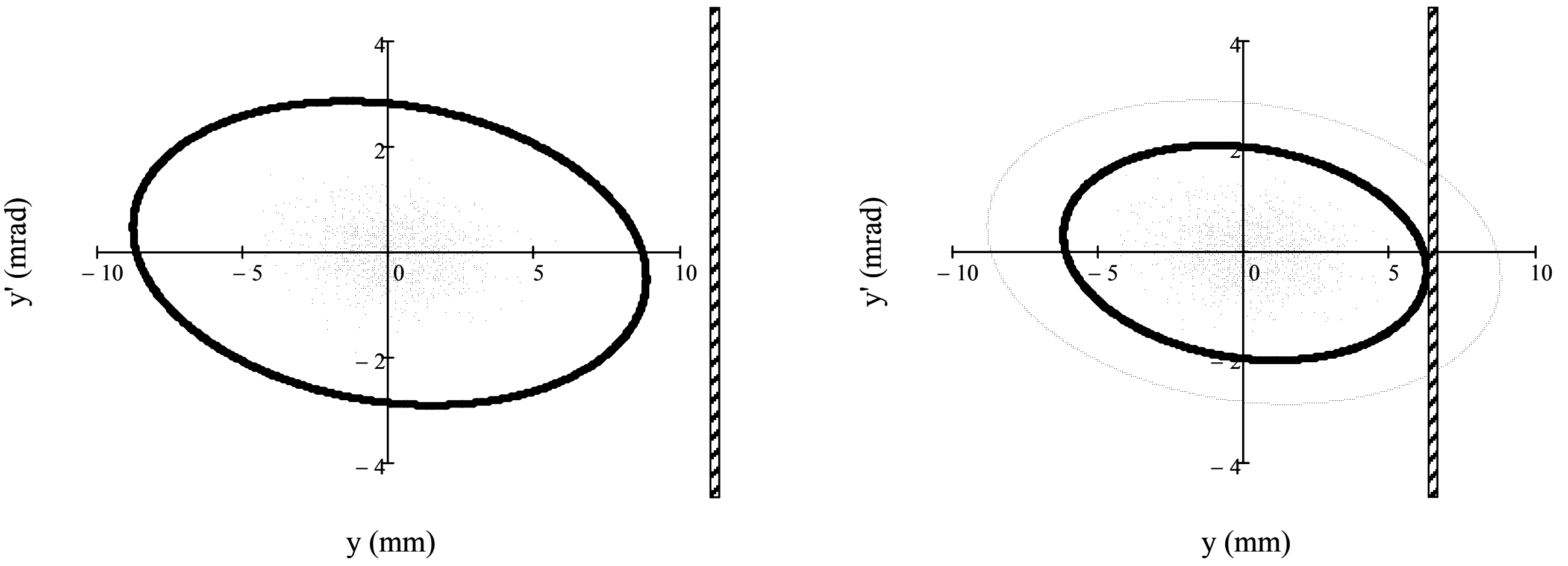

**Fig. 11.** y-y' phase space at the ANKE interaction point at 2.65 GeV with a machine acceptance of $A_y = 25\ \pi$ mm mrad, and a beam emittance of $\varepsilon_y = 3.8\ \pi$ mm mrad (left panel). When the aperture is moved into the machine acceptance (right panel), the reduced maximum allowed y' leads to smaller beam lifetime.

### 6.2 Machine acceptance at 2.65 GeV

The measured beam current as function of time with target on for different horizontal positions of the large diaphragm is plotted in Fig. 12. The main loss mechanism for beam particles is single Coulomb scattering on the target nuclei beyond the machine acceptance. Obviously the beam intensity can be approximated by an exponential only when the edge of the diaphragm is far away from the beam. One can identify two mechanisms that contribute to the non-exponential behavior shown in Fig. 12. Emittance growth widens the beam in proportion to the elapsed time. In addition, during the ten minute cycle, the beam is slowly decelerated because of the energy loss in the target. At ANKE, the horizontal dispersion is very small (see Fig. 6, bottom right panel). The large dispersion of $D_x \sim 15$ m in the arcs, however, together with the change in beam momentum $\Delta p$ due to the energy loss leads to a horizontal shift $\Delta x$ of the beam in the arcs, given by

$$\Delta x = D_x \cdot \frac{\Delta p}{p}. \tag{9}$$

In general, both mechanisms, emittance growth and transverse displacement of the beam, lead to an additional loss of beam particles *elsewhere* in the machine, i.e. not at the location of the large diaphragm, and cause the time dependence of the beam lifetime. During the measurements described here, the emittance growth was small, as discussed in ref. [3]. The Coulomb loss cross section is inversely proportional to the square of the acceptance angle x' at the limiting aperture [9]. Since, as discussed above, $x' \propto x$, it is easy to see that

$$\tau \propto x'^2 \propto x^2. \tag{10}$$

The additional beam loss due to the large dispersion in the arcs can therefore be accounted for by a correction term to the beam lifetime quadratic with respect to time, hence

$$\tau(t) = \tau_0 + a \cdot t^2, \tag{11}$$

and the beam intensity has been parameterized by the exponential function

$$N(t) = N_0 \cdot \exp\left(-\frac{t}{\tau(t)}\right). \tag{12}$$

The time-independent component of the beam lifetime $\tau_0$ corresponds to the beam lifetime one would observe when the mean energy loss is compensated by phase-space cooling, or by using an RF cavity with a bunched beam. A fitting procedure has been applied taking into account an absolute error of the beam current of $\Delta I = 1$ mV. Although not presented here graphically, the time-independent beam lifetimes $\tau_0$ from the vertical displacement of the diaphragm has been extracted in the same way.

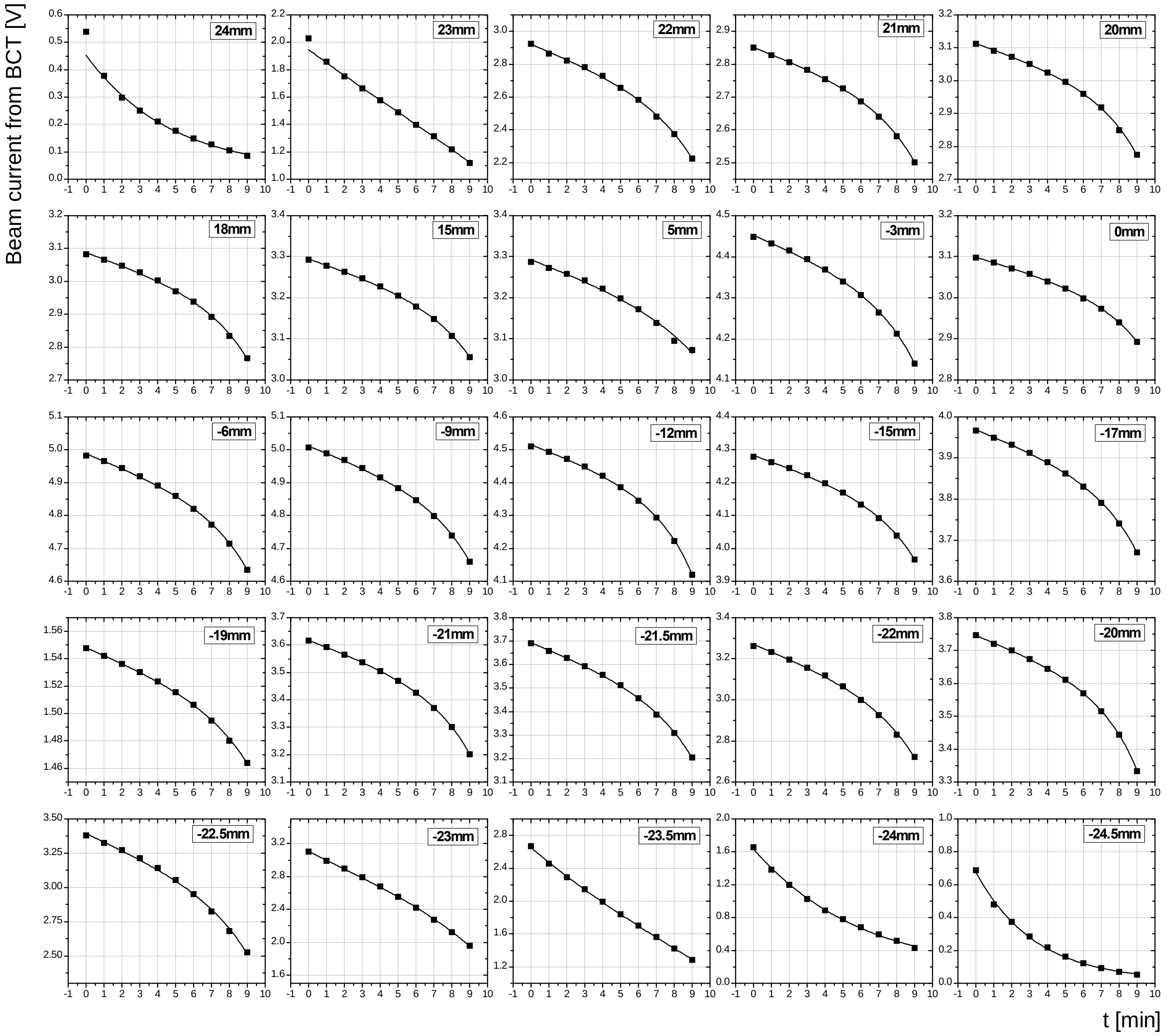

**Fig. 12.** Scan along the x axis with target switched on. Plotted is the beam current from the BCT (in V) versus time (in min). A fixed uncertainty of $\Delta I = 1$ mV is assumed. The lines indicate the results of a fit with Eq. (12).

In Fig. 13, $\sqrt{\tau_0}$ is plotted as a function of the horizontal (and vertical) displacement of the large diaphragm. As expected, as long as the aperture restriction is outside of the machine acceptance, the beam lifetimes remain constant. When the diaphragm cuts into the horizontal (or vertical) machine acceptance, the square root of the beam lifetime $\tau_0$ drops linearly with increasing displacement of the diaphragm $x$ (Eq. (10)). The result of a fit with a trapezoid-shaped model function is indicated in Fig. 13. The base of the approximated trapezoid for the vertical (horizontal) displacement corresponds to the height (width) of the large diaphragm of 25 mm (50 mm). It should be noted that this simple model function does not account for all physical effects involved. Therefore, in order to obtain estimates of the machine acceptance from the measurement, the uncertainties of $\sqrt{\tau_0}$ have been scaled by a factor to yield a reduced $\chi^2$ of approximately unity. The parameters of the trapezoid-shaped approximations are given in Fig. 13. Height (and width) of the free machine aperture at the ANKE interaction point are obtained by subtracting half the width of the plateau region from half the height (and width) of the large diaphragm. In Table 4, the results are listed together with the β functions, and the calculated machine acceptance.

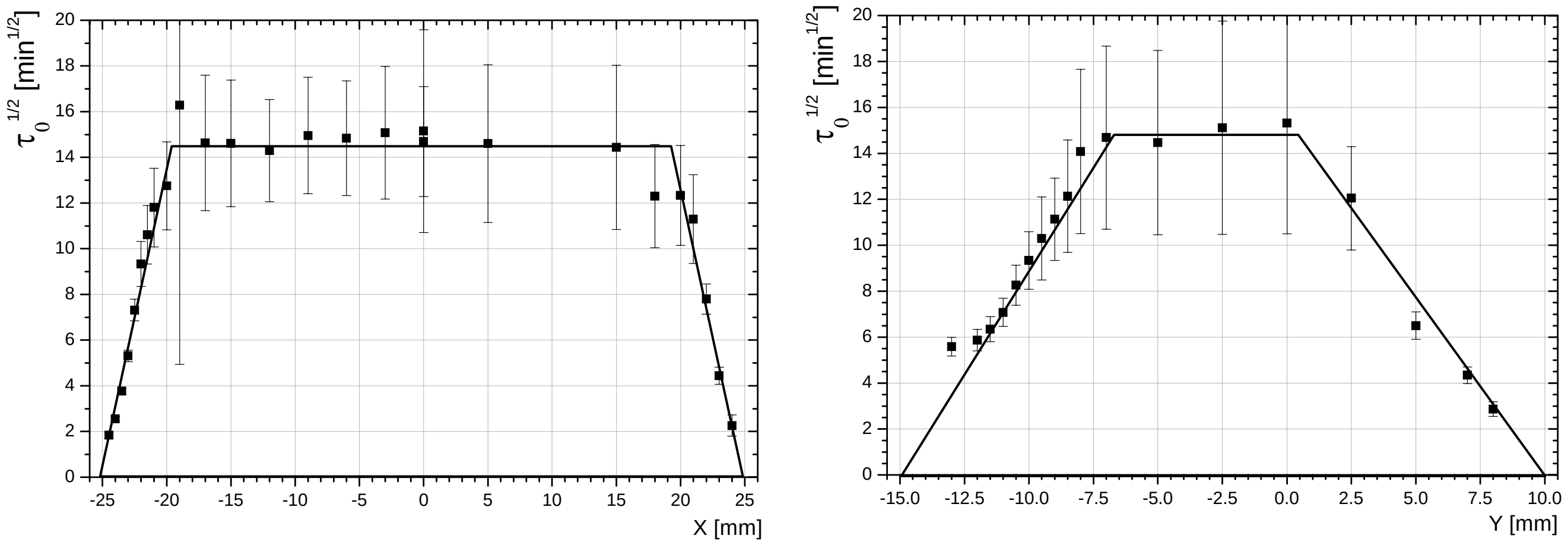

| | | |
|---|---|---|
| 19.24 ± 0.37 | half width of plateau [mm] | 3.67 ± 1.54 |
| 14.39 ± 0.83 | $\sqrt{\tau_0}$ in the plateau [$min^{1/2}$] | 14.91 ± 2.57 |
| -0.09 ± 0.04 | offset with respect to center [mm] | -3.06 ± 0.10 |

**Fig. 13.** Square root of the COSY beam lifetime with cluster target switched on at 2.65 GeV as function of the diaphragm shift in the horizontal (left panel) and vertical direction (right).

Comparing the results for the offsets of the beam from the center of the target chamber obtained from the beam-size measurements (Fig. 10) and from the analysis of the beam lifetimes (Fig. 13), one can conclude that the measured vertical beam offsets agree very well with each other. It should be noted that the horizontal shift of the beam position of Δx=+0.3 mm observed using 10 min long cycles could be caused by a small *negative* dispersion at the ANKE target position, contrary to the small positive one obtained from the lattice calculations, shown in Fig. 6.

**Table 4.** Measured free vertical and horizontal machine aperture at ANKE after acceleration to 2.65 GeV. Using the known β functions, the machine acceptance is calculated, taking into account a relative error of 10% for the β functions.

| | half width of free aperture $w$ | β function | Δβ/β | machine acceptance $w^2/\beta$ |
|---|---|---|---|---|
| | [mm] | [m] | | [π mm mrad] |
| x | 5.76 ± 0.37 | 2.6 | 0.1 | 12.74 ± 2.07 |
| y | 8.86 ± 1.53 | 3.1 | 0.1 | 25.33 ± 9.08 |

Measurements with the ANKE cluster target switched off were carried out as well. In this case, the beam lifetimes become much larger, reaching values of 120 ± 15 h, which is in good agreement with previously published results [17] at the highest COSY energies. The uncertainties involved, however, did not allow a reliable extraction of the free machine aperture without target, and hence a determination of the machine acceptance without target was not possible.

During the measurements with the small diaphragm ($10\ \text{mm}_{\text{ver}} \times 30\ \text{mm}_{\text{hor}}$), single injections showed that not more than 10% of the initial beam intensity could be stored, and after acceleration the beam was completely lost. Apparently, the orbit motion in the machine during acceleration requires that the vertical dimension of the storage cell tube should not be smaller than 15 mm. Thus, to be on the safe side for the upcoming experiments at ANKE, the storage cell should possess dimensions of $15\ \text{mm}_{\text{ver}} \times 20\ \text{mm}_{\text{hor}}$, where the horizontal cell size of 20 mm is chosen to ensure high intensity at injection (see Table 1).

## 7 Estimate of the luminosity with the polarized storage cell target

For the future measurements with the polarized internal gas target at ANKE, the ABS will be used to feed the storage cell with polarized hydrogen or deuterium atoms. In order to provide an estimate of the available luminosity for a prototype storage cell, a rectangular beam-tube with dimensions of $15\ \text{mm}_{\text{ver}} \times 20\ \text{mm}_{\text{hor}} \times 390\ \text{mm}_{\text{long}}$ and a feeding tube of length $L_3 = 150$ mm and diameter $D_3 = 15$ mm will be used. Space restrictions at the ANKE spectrometer require the use of an asymmetric cell with $L_1 = 160$ mm and $L_2 = 230$ mm (see Fig. 2). In this case, the total conductance of the cell is given by Eq. (4). The conductance of the beam tube of cross section a×b with round corners of radius r = 2 mm is approximated by the corresponding diameter of a round tube of $D_1 = D_2 = 2 \cdot \sqrt{\frac{a \times b - r^2(4-\pi)}{\pi}}$, yielding $C_{tot} = 46.4 \cdot 10^3\ \text{cm}^3/\text{s}$. The ABS delivers a H beam of $7.6 \cdot 10^{16}$ atoms/s in two hyperfine states [13]. Under these conditions, the estimated areal target density using Eqs. (3) and (5) reaches values around $d_t = 3.2 \cdot 10^{13}\ \text{atoms/cm}^2$. Taking into account a COSY beam intensity of $n = 10^{10}$ stored particles after acceleration to 2.65 GeV ($f_{rev} = 1.6$ MHz), the luminosity in polarized experiments should be about

$$L = \frac{1}{2} \cdot d_t \cdot n \cdot f_{rev} = \frac{3.2 \cdot 10^{13}}{2}\ \text{cm}^{-2} \times 10^{10} \times 1.6 \cdot 10^6\ \text{s}^{-1} = 2.5 \cdot 10^{29}\ \text{cm}^{-2}\text{s}^{-1}, \qquad (13)$$

where the factor ½ accounts for the fact that single hyperfine states will be injected into the cell.

## 8 Conclusion

The main goal of the investigations presented in this paper was to determine the optimum dimensions of a storage cell to provide high luminosity in internal experiments with the PIT at ANKE. A dedicated movable scraper system using diaphragms at the ANKE target location has been developed for that purpose. Using this system, the size of the COSY beam at 45 MeV (injection) and at 2.65 GeV was determined, from which the horizontal and vertical beam emittances are calculated.

At 2.65 GeV, the beam scraper system was also used to determine the machine acceptance, for which a novel technique was applied which is based on the measured beam lifetimes with a cluster target of target thickness ~$10^{14}$ atoms/$cm^2$. The dimensions for the storage cell were then optimized based on the beam sizes and machine acceptances measured in the target region, yielding 15 mm $_{ver}$ × 20 mm $_{hor}$ × 390 mm $_{long}$. A storage cell of these dimensions should provide a good compromise between beam intensity losses at injection and the desired increase in target density by the cell itself. It should be noted that the beam intensity at injection could be further increased by applying cooler stacking and stochastic cooling. The PIT at ANKE, equipped with such a cell, should provide luminosities for the experiments of more than $2.5 \cdot 10^{29}$ $cm^{-2}s^{-1}$ [7].

As a future upgrade option, we are presently developing a storage cell that can be opened during injection. Thereby, the beam losses at injection and from the transverse motion of the beam during acceleration can be avoided altogether. An *openable* cell, when closed, should have a cross section of 15 mm $_{ver}$ × 12 mm $_{hor}$, yielding an increase in luminosity compared to the one given in Eq. (13) by a factor of $(20/12)^3$~5.

## Acknowledgments

The authors would like to thank the members of the COSY crew and the ANKE collaboration for their support. We would also like to acknowledge the support by the COSY-FFE program (contract COSY-059) and the support by the BMBF through the contracts 06 ER 831 and 06 ER 126.